\documentclass{article}
\usepackage{amssymb}

\usepackage{graphicx}
\usepackage{amsmath}

\input{tcilatex}

\begin{document}

\title{Quantum tunneling and the principle of relativity. }
\author{\textbf{Elmir Dermendjiev} , \\
Institute for Nuclear Research and Nuclear Energy \\
of the Bulgarian Academy of Sciences, Tzarigradsko Shaussee, 72, \\
1784-Sofia, Bulgaria}
\maketitle

\begin{abstract}
It is shown that the results of Buttiker and Landauer on the traversal time
of quantum tunneling through a potential barrier are in agreement with the
principle of relativity. Also, they are consistent with the data on the
life-time of nuclear particles that decay in flight.

\end{abstract}

A tunneling of particles through a potential barrier is a quantum effect,
which has been intensively studied during the last decade.. Many authors
have studied the traversal time necessary the barrier to be passed, however
the aim of this letter is to discuss the results obtained by Buttiker and
Landauer [1]. Landauer and Martin [2] considered different points of view
[3,4] on the nature of the traversal time. Due to their analysis, it was
shown that the Buttiker-Landauer traversal time $\tau _{BL}$ seams to
correspond quite well to the interaction time between the tunneling paticle
and the barrier and can be expressed by the following equation [1]:

\begin{equation}
\tau _{BL}=\int_{x_{1}}^{x_{2}}\frac{m}{\hbar k(x)}dx=\int_{x_{1}}^{x_{2}}%
\sqrt{\frac{m}{2\hbar \lbrack V_{0}(x)-E]}}dx
\end{equation}

Here,$m$ and $E$ are the mass and the energy of the particle, $V_{0}(x)$-is
the rectangular potential barrier with turning points $x_{1}$ and $x_{2}$.
The eq. (1) is valid if $V_{0}(x)>>E$ [1].

An attempt to estimate the value of the $V_{0}(x)$ for strong interacting
particles has been done in [5,6]. It was shown that the value of such
barrier seams to be very high, since $V_{0}^{si}(x)$'' 10$^{20}m$ [6], where
m is the mass of a particle. If so, then one can assume that the barrier can
be passed when $E=\frac{mv_{b}^{2}}{2}$. It is expected that the velocity vb
is very high. However, to satisfy the principle of relativity it is
necessary that always $v_{b}<C$, where $C$ is the velocity of light. If $%
v_{b}->$ $C$, then one can transform the eq.(1) to the following equation:

\begin{equation}
\tau _{BL}=\frac{\Delta x}{C\sqrt{1-\frac{v^{2}}{C^{2}}}}
\end{equation}

No doubts that the eq. (2) has fundamental significance. However, due to the
lack of space we limit comments of eq. (2) giving only one clear example.
Since the eq. (1) contains a relativistic multiplier, the decay in flight of
particles and their larger life-time $\tau $ relative to the life-time in
rest $\tau _{0}$ are consistent with the principle of relativity. Assuming
that $\tau _{BL}\thickapprox $ $\tau $ one can reach the following simple
relationship: 
\begin{equation}
\tau _{0}=\frac{\Delta x}{C}
\end{equation}

The eq. (3) shows that the value of the width $\Delta x$ of the potential
barrier is equal to the ''characteristic length'' ($\tau _{0}C$) of
elementary particles. Estimated values of $\Delta x$ for $\tau _{0}$ $%
\thickapprox $\ $2,2.10^{-6}$ s ($\mu ^{\pm }$ - mesons); $2,6.10^{-8}s$ ($%
\pi ^{\pm }$ - mesons) and $0,8.10^{-16}$s ($\pi ^{0}$-meson) are as follow: 
$\thickapprox \ 6,6.10^{4}cm$, \ $\thickapprox 7,8.10^{2}cm$ and \ $%
\thickapprox 2,4.10^{-6}cm$. These surprising results lead us to the
conclusion that the decay in flight of particles could have been associated
with a tunneling of particles through a potential barrier.

If the potential barrier is time-dependent, i.e. if the barrier $%
V(x,t)=V_{0}(x)+V_{1}(x)Cos(\omega t)$[1], then one can find some other
important consequences from the results of Buttiker and Landauer's work [1].

To derive them, one should consider the results and numerical evaluations
obtained in [5,6]. It was assumed that the time associated with a given
particle or body depends from the relative changes of their masses, i.e. - $%
mdt=\frac{\Delta m}{m}$ [5,6]. An estimation of the value of $\mu _{n}$ for
strong interacting particles was found to be $\thickapprox 10^{23}s^{-1}$%
[5]. A phenomenological approach based on the assumption that the strong
interacting field ''vibrates'' with a frequency mn was considered in [5,6].
One could assume that the potential barrier $V(x,t)$ [1] is modulated with a
frequency $\varpi $ $\thickapprox \mu _{n}$. Then the tunneling particles
are expected to have energies with values of $E\pm n\hbar \varpi $ that form
two ''sidebands'' [1]. An estimation of the change of $E$ at $n=1$ gives a
value of $\Delta E$ $\thickapprox \hbar $.$\mu _{n}$ $\thickapprox $ $66MeV$%
. It is important to note that both charged and neutral particles are
expected to change their energies. This important result might be a physical
basis for acceleration of strong interacting particles that are tunneling
through a time-dependent, i.e. modulated, potential barrier.

One should note that an experimental study of the energy spectra of
tunneling nuclear particles that pass through a modulated potential barrier
could check all assumptions mentioned above. Also, one could obtain unique
information about some of the properties of strongly interacting field.

In conclusion, one should note that it would be interesting further
consideration of some of the properties of nuclear matter to be done from
the point of view of quantum tunneling with dissipation [1] and a resonant
interaction between particles and barriers [7].

References

[1] M.Buttiker and R.Landauer, Phys.Rev.Letters 49(1982)1739

[2] R.Landauer, Th.Martin, Rev.Mod.Phys. 66(1994)217

[3] E.P.Wigner, Phys.Rev. 98(1955)145

[4] C.R.Leavens and G.C.Aers, in ''Scanning Tunneling Microscopy''

1993, Springer, New York, p.105

[5] E.Dermendjiev, Report nucl-th 0011028 (8 November 2000)

[6] E.Dermendjiev, Report nucl-th 0011035 (9 November 2000)

[7] M.Ya.Azbel and V.M.Tzukernik, Europhys.Lett., 41(1998)7

\end{document}